# Diagnosing Renyi and Tsallis Holographic Dark Energy Models with Hubble's Horizon Cutoff

A.Y.Shaikh*

*Department of Mathematics, Indira Gandhi Mahavidyalaya , Ralegaon 445402, India.

e-mail: shaikh_2324ay@yahoo.com

**Abstract:**

In this work, I framed the Renyi and Tsallis Holographic Dark Energy (HDE) models within the presence of a spatially flat and isotropic FRW model filled with matter and dark energy in teleparallel gravity. The energy densities of Renyi HDE and Tsallis holographic dark energy (THDE) are increasing functions of $z$ and validate the expanding conduct of the universe. It is worthwhile to note that for $-1 \leq z \leq 0$, the EoS parameter of Renyi HDE approaches to phantom region , while THDE confirms to Quintessence region. Further for $0 < z$, the EoS parameter of Renyi HDE slants to $\Lambda$ CDM model ,while THDE settles to Quintessence region. The model behaves like $\Lambda$ CDM with the statefinder parameters having the values $\{1,0\}$.

## 1. Introduction

Supernovae type Ia [1-2] and cosmic microwave background radiation [3-4] investigational verified the current fast-tracked expansion of the cosmos. There are two ways for describing an accelerating expansion of the cosmos: a) An enigmatic force with enormous negative pressure, named Dark Energy (DE) (see [5-8] and references therein), and b) modification of general theory of relativity [9-15]. In Ref. [16], the authors recommended by the WMAP experiment that the universe is composed of 73% DE, 23% dark matter, and 4% baryonic matter. There are different kinds of DE models which are categorized as quintessence [17], k-esssence [18], Chaplygin gas [19], holographic dark energy [20-21], new agegraphic dark energy [22], etc. In [23], the current observations suggest cosmological constant i.e. $\omega \approx -1$.

In modern eras, the research of holographic dark energy model turns out to be a promising method to elucidate the cosmic expansion [24] in the contextual of holographic principle (HP) [20-21]. Other readings are also accessible (check references for illustration [25-42]). The HDE energy density is written off as as $\rho_{de} = 3c^2 M_{pl}^2 L^{-2}$ reliant on the entropy-area relation of black holes [24], where $c$ is a numerical constant. A correlation amongst system entropy ($S$), the IR ($L$) and UV ($\Lambda$) cutoffs is framed as $L^3 \Lambda^3 \leq S^{\frac{3}{4}}$ [24]. After this work in [43], the authors have presented $S_\delta = \gamma A^\delta$ (the horizon entropy of a black hole), where i) $A$ is the area of Horizon, ii) $\delta$ corresponds to the non-additivity parameter, and iii) $\gamma$ redirects an unknown constant. This leads to $\Lambda^4 \leq (\gamma(4\pi)^\delta) L^{2\delta-4}$ [24]. In literature, Tsallis HDE (THDE) [43-44], Sharma-Mittal HDE (SMHDE) [45] and Renyl HDE model [46] are the new HDE models that have been constructed. In [47], the Renyi HDE model was examined, with the IR cutoff considered as the Hubble horizon. In Ref. [48], the authors have discussed Tsallis, Renyi, and Sharma-Mittal entropies in the context of Chern-Simons modified gravity while THDE in various braneworlds have been considered by Ghaffari et al. [49]. In [50], Jawad et al. have inspected Tsallis, Renyi, and Sharma-Mittal HDE models in loop quantum cosmology. In [51], Ghaffari et. al. studied the consequences of employing the THDE model in modeling dark energy in the BD cosmology. In the work of [52], the authors examined the interacting THDE model from the statefinder point of view and concluded that the interacting THDE can have a significant effect on the cosmic evolution of the universe. Recently in Yadav [53], the energy conservation law for THDE does not hold in Brans-Dicke gravity is observed.

Modified theories of gravity, such as $f(R)$ [54], $f(T)$ [55], $f(R,T)$ [56], and $f(G)$ [57], are desired to witness accelerating expansion and to provide an alternate for the DE. In [58-61], the torsion term $T$ in the teleparallel scenario is altered from the curvature term $R$ in general relativity, which alters $T$ to $f(T)$ by an arbitrary function with a varying action known as $f(T)$ gravity. Bamba and Geng explored the thermodynamics for apparent horizon in [62]. Charged wormhole solutions gravity with non-commutative background has been extensively explored in [63]. The dynamical instability of a spherically symmetric collapsing star is analyzed in [63]. Many researchers have worked on $f(T)$ gravity in recent past on different aspects of cosmology [64-81].

The main goal for this work is to illuminate the cosmic expansion in spatially homogeneous and isotropic flat FRW universe by using RHDE and THDE SMHDE and taking Hubble's Horizon Cutoff, which has not been explored earlier. An astrophysical scale factor in the form of power law is considered. Motivated by the above discussed recent works of various authors, I propose to investigate the Renyi and Tsallis HDE models within the context of theory of $f(T)$ gravity.

## 2. Analysis of $f(T)$ cosmology

The action of teleparallel gravity in the framework of $f(T)$ theory is given in [62] as

$$S = \int [T + f(T) + L_{matter}] e \, d^4x. \tag{1}$$

Here $f(T)$ denotes an algebraic function of the torsion scalar $T$. Taking the variation of the action (1) with respect to the vierbein in [82-83], one can obtain the field equations as

$$S_\mu^{\nu\rho} \partial_\rho T f_{TT} + [e^{-1} e_\mu^i \partial_\rho (e e_i^\alpha S_\alpha^{\nu\rho}) + T^\alpha_{\lambda\mu} S_\alpha^{\nu\lambda}](1 + f_T) + \frac{1}{4} \delta_\mu^\nu (T + f) = T_\mu^\nu, \tag{2}$$

where the energy momentum tensor is $T_\mu^\nu$, $f_T = \frac{df(T)}{dT}$. The energy momentum tensor for matter and DE is given by $T_\mu^\nu = T_{\mu\nu}^m + T_{\mu\nu}^{de}$, where $T_{\mu\nu}^m = \rho_m u_\mu u_\nu$; $T^{de}_{\mu\nu} = (\rho_{de} + p_{de}) u_\mu u_\nu - g_{\mu\nu} p_{de}$, where $\rho_m$ and $\rho_{de}$ illuminate the energy density of matter and DE density. $p_{de}$ is the pressure of the DE while equation of state (EoS) is defined as $\omega_{de} = \frac{p_{de}}{\rho_{de}}$. The energy-momentum tensor of DE can be parameterized as $T_{\mu\nu}^{de} = (-\omega_{de}, -\omega_{de}, -\omega_{de}, 1) \rho_{de}$, \tag{3} where $\omega_\tau$ are the directional EoS parameters on $x$, $y$ and $z$ axis, respectively.

## 3. Metric and the field equations

The FRW line element represented by the following metric

$$ds^2 = dt^2 - a^2(t) \left\{ \frac{dr^2}{1 - kr^2} + r^2 (d\theta^2 + \sin^2\theta \, d\phi^2) \right\}, \tag{4}$$

where the $0 \leq \theta \leq \pi$ and $0 \leq \phi \leq \pi$ are the azimuthal and polar angles of the spherical co-ordinate system. With respect to the curvature of the space represented by $k$, one can have the following points to be well thought-out. i) $k=1$ corresponds to closed universe, ii) $k=0$ agrees to flat universe and iii) $k=-1$ relates to open universe. In this work, the flat universe i.e. $k=0$ is deliberately considered. The torsion scalar is obtained as $T = -6H^2$, where $H = \frac{\dot{a}}{a}$ is the Hubble parameter and $a$ is the scale factor. For the line element (4), using equation (3), the field equations (2) takes the following form

$$\frac{\dot{a}}{a}\dot{T}f_{TT} + \left[\frac{\ddot{a}}{a} + 2\frac{\dot{a}^2}{a^2}\right](1+f_T) + \frac{1}{4}(T+f) = -\omega_{de}\rho_{de}, \tag{5}$$

$$3(1+f_T)\frac{\dot{a}^2}{a^2} + \frac{1}{4}(T+f) = (\rho_m + \rho_{de}), \tag{6}$$

where the overhead dot denotes the differentiation with respect to time.

### 4. Solutions of the field equations

It is observed that there are six unknowns with two field equations (5) and (6). In order to obtain the exact solutions, one can assume a state equivalent to some physical circumstances or an arbitrary mathematical hypothesis. Let us consider the scale factor of the form $a = t^n$ [84-86], where $n > 0$ is constant. The motivation to choose such scale factor is behind the fact that the universe is accelerated expansion at present and decelerated expansion in the past. For the diverse values of $n$, different stages of the universe are prophesied. For $0 < n < 1$, the model corresponds to the decelerating phase whereas the universe accelerates for $n > 1$. In the case of $n = 1$, the value of deceleration parameter i.e. $q = 0$ implies the universe inflates and the average scale factor has a linear growth with constant velocity. The model exhibits singularity at an initial epoch. Hence at $t \to 0$, the model starts evolving with a big-bang. The scale factor $a(t)$ and the redshift $z$ are associated through the relation $(1+z) = \frac{a_0}{a}$, where $a_0$ represent the present value of scale factor. The cosmological parameter called as deceleration parameter, which measures the rate of expansion of the universe is defined and calculated as $q = \frac{d}{dt}\left(\frac{1}{H}\right) - 1 = \frac{1}{n} - 1$. The deceleration parameter shows signature flipping on the constraints of $n$. The universe inflates for $q < 0$, while it decelerates for $q > 0$. The deceleration parameter attains $q \cong -0.733$ after choosing the value $n = 3.75$. According to the $\Lambda$CDM observational data the current value of the deceleration parameter may be $q_0 = -0.6$.

### 5. Renyi Holographic Dark Energy Model with Hubble's Horizon Cutoff

The energy density of Renyi HDE is $\rho_{de} = \frac{3\beta^2}{8\pi L^2}\left(1 + \pi\delta L^2\right)^{-1}$, where $\beta$ and $\delta$ are constants. By considering the Hubble horizon as the IR cutoff of the system, i.e., $L = H^{-1}$, the Renyi HDE density is

$$\rho_{de} = \frac{3\beta^2 n^2}{8\pi t^2}\left(1 + \frac{\pi\delta t^2}{n^2}\right)^{-1}. \tag{7}$$

The energy density of matter is

$$\rho_m = \frac{3n^2}{2t^2} - \frac{3\beta^2 n^2}{8\pi t^2}\left(1 + \frac{\pi\delta t^2}{n^2}\right)^{-1}. \tag{8}$$

The equation of state parameter (the relationship between pressure $p_{de}$ and energy density $\rho_{de}$ of DE) is extensively used to classify the several segments of the escalating universe. It categorizes the decelerated and accelerated phases of the escalating universe in the below mentioned eras. If $\omega = 0$ then it symbolizes cold dark matter or dust fluid, $\omega = 1$ yields stiff fluid and radiation era for $0 < \omega < \frac{1}{3}$. The model illustrates the quintessence phase when $-1 < \omega < -\frac{1}{3}$. It yields the cosmological constant, i.e., $\Lambda$ CDM model for $\omega = -1$ and $\omega < -1$ signifies the phantom era. Quintom era is the combination of both quintessence and phantom. The EoS parameter of Renyi HDE is obtained as

$$\omega_{de} = \frac{\dfrac{-n^2}{t^2}}{\dfrac{3\beta^2 n^2}{8\pi t^2}\left(1+\dfrac{\pi\delta t^2}{n^2}\right)^{-1}}. \qquad (9)$$

The density parameter of Renyi HDE and the energy density parameter of matter are defined and obtained as

$$\Omega_{de} = \frac{\rho_{de}}{3H^2} = \frac{\left\{\dfrac{3\beta^2 n^2}{8\pi t^2}\left(1+\dfrac{\pi\delta t^2}{n^2}\right)^{-1}\right\}}{\dfrac{3n^2}{t^2}}, \quad \Omega_m = \frac{\rho_m}{3H^2} = \frac{\left\{\dfrac{3n^2}{2t^2}-\dfrac{3\beta^2 n^2}{8\pi t^2}\left(1+\dfrac{\pi\delta t^2}{n^2}\right)^{-1}\right\}}{\dfrac{3n^2}{t^2}}. \qquad (10)$$

Hence, $\Omega = \Omega_m + \Omega_{de} = \dfrac{\left\{\dfrac{3\beta^2 n^2}{8\pi t^2}\left(1+\dfrac{\pi\delta t^2}{n^2}\right)^{-1}\right\}}{\dfrac{3n^2}{t^2}} + \dfrac{\left\{\dfrac{3n^2}{2t^2}-\dfrac{3\beta^2 n^2}{8\pi t^2}\left(1+\dfrac{\pi\delta t^2}{n^2}\right)^{-1}\right\}}{\dfrac{3n^2}{t^2}}. \qquad (11)$

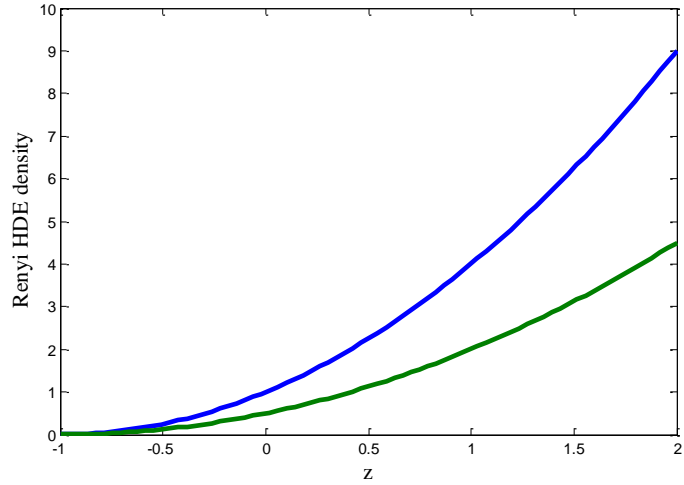

Figure 1. Plot of Renyi HDE density against z for $\beta = 1.2, \delta = 4, n = 1.5$ (Blue) and

$\beta = 1.2, \delta = 4.5, n = 1.5$ ( Green).

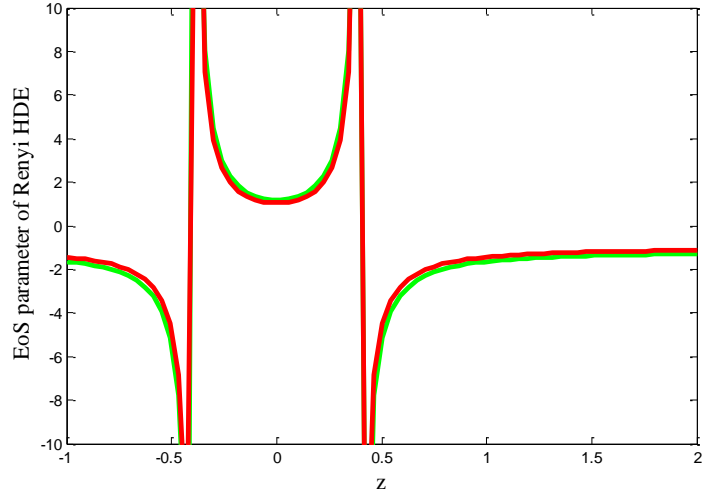

Figure 2. Plot of EoS parameter of Renyi HDE against z for $\beta = 1.2, \delta = 4, n = 1.5$ (Red) and

$\beta = 1.2, \delta = 4.5, n = 1.5$ ( Green).

Figure 1 portrays the conduct of Renyi HDE density with the Hubble horizon cutoff versus $z$ for the different values of $\delta = 4$ and $\delta = 4.5$. It is observed that $\rho_{de}$ is positive and decreases with the evolution of the universe. It can be seen that the energy density of matter decreases from a high red-shift region to the low red-shift region i.e. from past to future. The behavior of EoS parameter of Renyi HDE with the Hubble horizon cutoff versus $z$ for the different values of $\delta = 4$ and $\delta = 4.5$ is depicted in figure 2. For all $-1 \leq z \leq 0$, it is witnessed that the model starts from matter dominated era, crosses quintessence phase $(-1 < \omega_{de} < -0.33)$, Cosmological constant i.e. $\Lambda$CDM model $(\omega_{de} = -1)$, and finally approaches to phantom region $(\omega_{de} < -1)$. Further, for $z > 0$, the EoS parameter of Renyi HDE slants to $\Lambda$CDM model. It is observed that the density parameter of Renyi HDE and matter become singular when $H = 0$. The total energy density approaches to a constant value which confirms the isotropic nature of the Universe.

## 6. Tsallis Holographic Dark Energy (THDE) Model with Hubble's Horizon Cutoff

The authors in the reference [43] shown that the Tsallis generalized entropy-area relation is independent of the gravitational theory used to study the system, hence the energy density of Tsallis holographic dark energy (THDE) is

$$\rho_{de} = L^{2\delta-4}, \tag{12}$$

where $\delta$ is constant. For the value of $\delta = 1$, the energy density of Tsallis holographic dark energy moderates to the energy density of HDE model. By considering the Hubble horizon as a candidate for the IR-cutoff, i.e., $L = H^{-1}$, which implies $\rho_{de} = \dfrac{1}{H^{2\delta-4}}$. Thus, the Tsallis holographic dark energy density is $\rho_{de} = \left(\dfrac{t}{n}\right)^{2\delta-4}$. \hfill (13)

The energy density of matter is $\rho_m = \dfrac{3n^2}{2t^2} - \left(\dfrac{t}{n}\right)^{2\delta-4}$. \hfill (14)

The EoS parameter of THDE is obtained as $\omega_{de} = \dfrac{\dfrac{-n^2}{t^2}}{\left(\dfrac{t}{n}\right)^{2\delta-4}}$ .(15)

The density parameter of THDE and the energy density parameter of matter are defined and obtained as

$$\Omega_{de} = \frac{\rho_{de}}{3H^2} = \frac{\left\{\left(\dfrac{t}{n}\right)^{2\delta-4}\right\}}{\dfrac{3n^2}{t^2}}, \Omega_m = \frac{\rho_m}{3H^2} = \frac{\left\{\dfrac{3n^2}{2t^2} - \left(\dfrac{t}{n}\right)^{2\delta-4}\right\}}{\dfrac{3n^2}{t^2}}. \quad (16)$$

Hence , $\Omega = \Omega_m + \Omega_{de} = \dfrac{\left\{\left(\dfrac{t}{n}\right)^{2\delta-4}\right\}}{\dfrac{3n^2}{t^2}} + \dfrac{\left\{\dfrac{3n^2}{2t^2} - \left(\dfrac{t}{n}\right)^{2\delta-4}\right\}}{\dfrac{3n^2}{t^2}}$ . (17)

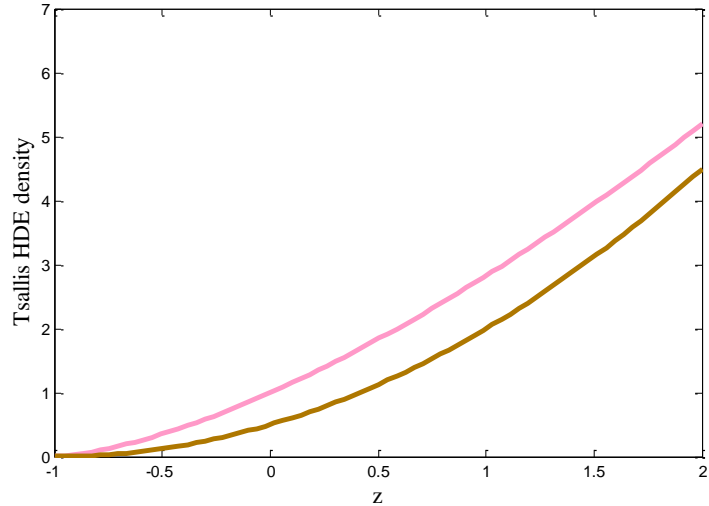

Figure 3. Plot of THDE density against z for $\delta = 4, n = 1.5$ (Brown) and $\delta = 4.5, n = 1.5$ ( Pink).

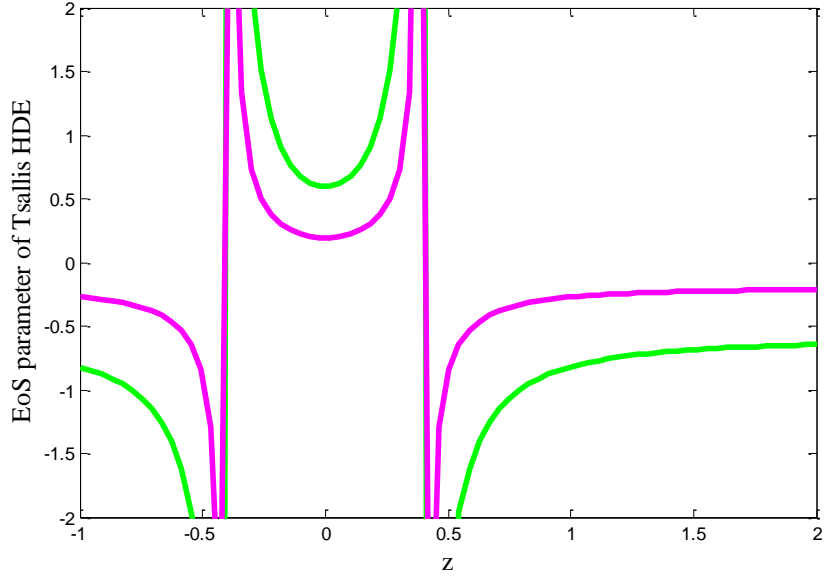

Figure 4. Plot of EoS parameter of THDE against z for $\delta = 4$, $n = 1.5$ (Green) and

$\delta = 4.5$, $n = 1.5$ (Pink).

Figure 3 represents the actions of THDE density with the Hubble horizon cutoff versus *z* for the different values of $\delta = 4$ and $\delta = 4.5$. It is observed that $\rho_{de}$ is positive and decreases with the evolution of the universe. It can be seen that the energy density of matter decreases from a high red-shift region to the low red-shift region i.e. from past to future. The behavior of EoS parameter of THDE with the Hubble horizon cutoff versus *z* for the different values of $\delta = 4$ and $\delta = 4.5$ is depicted in figure 4. For all $-1 \leq z \leq 0$, it is witnessed that the model starts from matter dominated era, crosses quintessence phase ($-1 < \omega_{de} < -0.33$), Cosmological constant i.e. $\Lambda$CDM model $(\omega_{de} = -1)$, Phantom divide line (PDL) and finally approaches to the Quintessence region. Further, for $z > 0$, the EoS parameter of THDE inclines to the Quintessence model. The total energy density predicts a flat universe as it tends to constant value for sufficiently large time.

# 7. Statefinder parameters

In [55], Sahni et al. introduced the cosmological diagnostic pair $\{r,s\}$, where $r$ and $s$ are defined

as $r = \dfrac{\dddot{a}}{aH^3}$ , $s = \dfrac{(r-1)}{3\left(q-\dfrac{1}{2}\right)}$ . \hfill (18)

The relation between the diagnostic parameters is obtained as $r = \dfrac{9s^2 - 9s + 2}{2}$ . \hfill (19)

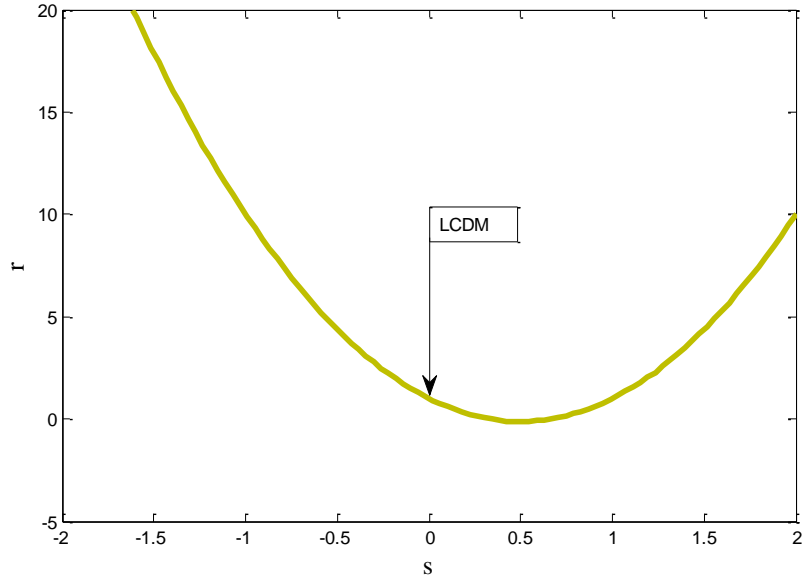

Figure 5. Plot of $r$ versus $s$.

The statefinder parameters are characterized as: i) for $(r,s) = (1,0)$, the model represents the $\Lambda$ CDM model, ii) for $(r,s) = (1,1)$, it yields SCDM model, iii) for $s > 0$ and $r < 1$, the model corresponds Quintessence region , iv) for $s < 0$ and $r > 1$, it indicates the Chaplygin gas , and v) for $(r,s) = \left(1, \dfrac{2}{3}\right)$, the model shows HDE . The model behavior is predicted in figure 5 which is similar to [87].

# 8. Observational Constraints

## Look back time redshift, Luminosity distance and Distance modulus

Keeping the relation $1+z = \dfrac{a_0}{a}$ in mind between the scale factor $a$ and redshift $z$, one can obtain

$1+z = \left(\dfrac{t_0}{t}\right)^n$, for $n \neq 0$. After mathematical manipulation, the above equation yields

$t = t_0 (1+z)^{-\frac{1}{n}}$. This equation can also be written as $H_0(t_0 - t) = n\left[1 - (1+z)^{-\frac{1}{n}}\right]$, where $H_0$ is the present value of the Hubble's parameter. For the small value of redshift $z$, above equation reduces to $H_0(t_0 - t) = n\left[\dfrac{z}{n} - \dfrac{\frac{1}{n}\left(\frac{1}{n}-1\right)}{2} z^2 + \ldots\ldots\right]$. Hence, $H_0(t_0 - t) = \left[z - \dfrac{q}{2n} z^2 + \ldots\ldots\right]$.

The distance modulus is expressed as $\mu(z) = 5\log d_L(z) + 25$, \hfill (20)

where $d_L(z)$ is the luminosity distance and is defined as $d_L = r_1(1+z)a_0$. For the determination of $r_1$, let us assuming that a photon emitted by source with coordinate $r = r_0$ to $t = t_0$ and received at a time $t$ by an observer located at $r = 0$, then one can determine $r_1$ from the relation

$$r_1 = \int_t^{t_0} \dfrac{dt}{a} = \int_t^{t_0} t^{-n} dt. \qquad (21)$$

Hence, the expression for luminosity distance can be obtained as

$$d_L = \dfrac{(1+z)H_0^{-1}[1-(1+z)^{1-\frac{1}{n}}]}{(1-n)}. \qquad (22)$$

It is observed that the luminosity distance increases faster with red shift, exactly as required by the supernova data. Using equations (20) and (22), it yields

$$\mu(z) = 5\log\left\{\frac{(1+z)H_0^{-1}[1-(1+z)^{1-\frac{1}{n}}]}{(1-n)}\right\} + 25. \tag{23}$$

The distance modulus of derived model is in good agreement with SN Ia data.

**Table 1.: Comparison between the Supernovae Ia data and present power law model**

| Redshift(z) | Supernovae Ia $(\mu)$ | Power law model $(\mu)$ | Redshift(z) | Supernovae Ia $(\mu)$ | Power law model $(\mu)$ | Redshift(z) | Supernovae Ia $(\mu)$ | Power law model $(\mu)$ |
|---|---|---|---|---|---|---|---|---|
| 0.014 | 33.73 | 33.79 | 0.240 | 40.68 | 40.72 | 0.930 | 44.61 | 44.70 |
| 0.026 | 35.62 | 35.66 | 0.380 | 42.02 | 42.20 | 0.949 | 43.99 | 43.99 |
| 0.036 | 36.39 | 36.42 | 0.430 | 42.33 | 42.37 | 0.970 | 44.13 | 44.17 |
| 0.040 | 36.38 | 36.54 | 0.480 | 42.37 | 42.51 | 0.983 | 44.13 | 44.17 |
| 0.050 | 37.08 | 37.15 | 0.620 | 43.11 | 43.17 | 1.056 | 44.25 | 44.28 |
| 0.060 | 37.67 | 37.78 | 0.740 | 43.35 | 43.46 | 1.190 | 44.19 | 44.20 |
| 0.079 | 37.94 | 37.99 | 0.778 | 43.81 | 43.88 | 1.305 | 44.51 | 44.56 |
| 0.088 | 38.07 | 38.06 | 0.828 | 43.59 | 43.67 | 1.340 | 44.92 | 44.97 |
| 0.101 | 38.73 | 38.78 | 0.886 | 43.91 | 43.99 | 1.551 | 45.07 | 45.10 |
| 0.160 | 39.08 | 39.10 | 0.910 | 44.44 | 44.52 | | | |

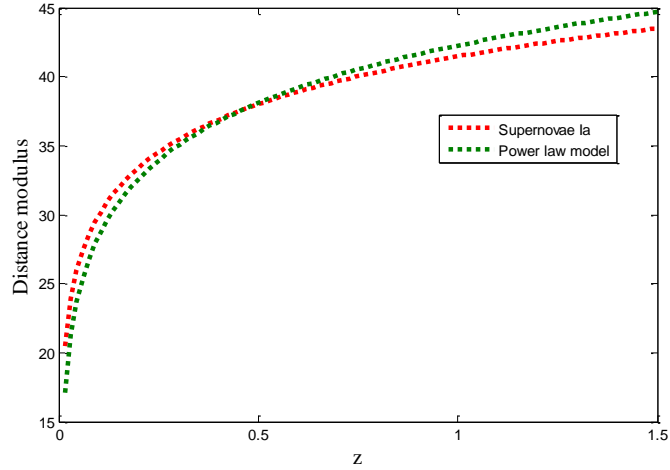

Figure 6: Distance modulus against z.

For the interval $0.014 < z < 1.551$, the data groups of supernova Ia in the choice are examined. The assessment concerning distance modulus $\mu$ of the power law model and the observational $\mu(z)$ SNe Ia data can be implicit in Table 1. It is perceived that the derived model is fit well with SNe Ia observation (Table 1) which are physically realistic.

## 9. Discussion and concluding remarks

In the present work, Renyi and Tsallis HDE models have been chosen within the context of $f(T)$ theory of gravity. It is observed that the energy densities of Renyi HDE and THDE with the Hubble horizon cutoff are increasing functions of $z$ and validate the expanding conduct of the universe. For $-1 \leq z \leq 0$, the EoS parameter of Renyi HDE approaches to phantom region, while THDE confirms to Quintessence region. Further for $0 < z$, the EoS parameter of Renyi HDE slants to $\Lambda$CDM model, while THDE settles to Quintessence region. Hence, the EoS parameters of Renyi HDE and THDE lie in the accelerated stage dominated by DE era and deliver the consistency of the model. The derived model behaves like $\Lambda$CDM model for $(r,s) = (1,0)$ which resembles with [88-90]. As a result, the resultant models display reliability through the recognized theoretical outcomes as well as the current remarks of the cosmic expansion.